\documentstyle [11pt]{article}

\begin{document}

\begin{center}
{\Large\bf 
Single-mode realizations of the Higgs algebra
}
\end{center}

\vspace{.3cm}

\begin{center}
Dong Ruan
\end{center}
Department of Physics, Tsinghua University, 
      Beijing 100084, People's Republic of China, 
Key Laboratory for Quantum Information and 
	Measurements of MOE, Tsinghua University, 
	Beijing 100084, People's Republic of China and 
Center of Theoretical Nuclear Physics, 
      National Laboratory of Heavy Ion Accelerator, 
      Lanzhou, 730000, People's Republic of China

\vfill

\hfill     {Typeset by La\TeX}

\newpage

\begin{abstract}
In this paper we obtained for the Higgs algebra three kinds of single-mode realizations
such as the unitary Holstein-Primakoff-like realization, the non-unitary Dyson-like 
realization and the unitary realization based upon Villain-like realization. 
The corresponding similarity transformations between the Holstein-Primakoff-like 
realizations and the Dyson-like realizations are revealed.
\end{abstract}

\vspace{.5cm} 



\newpage

In recent years, polynomial angular momentum algebra (PAMA) and its
increasing applications in quantum problems have been the focus of 
very active research.
This kind of PAMA, spanned by three elements $J_{\mu}$ 
($\mu = +$, $-$, $3$), 
has a coset structure $h+ v$ \cite{rocek}, where $h$ is an ordinary 
Lie algebra U(1) generated by $J_3$; the remaining two elements 
$J_+$, $J_-$ $\in v$ transform according to a representation of U(1), 
and their commutator yields a polynomial function of $J_3 \in$ U(1).  
Hence, PAMA can be viewed as a deformation of an ordinary 
angular momentum algebra SU(2). \cite{bl}

In fact, the first special case of PAMA is the so-called Higgs algebra,
here denoted by ${\cal H}$.
In 1979, Higgs \cite {higgs} found that there exists a kind of PAMA with an extra 
cubic term in the isotropic oscillator and Kepler potentials in 
a two-dimensional curved space. 
Zhedanov \cite{zhedanov} presented a connection between the Higgs algebra
${\cal H}$ and the quantum group SU$_q$(2) \cite{jimbo}. 
Daskaloyannis \cite{dask} and Bonatsos \cite{bdk,bkd} discussed the PAMA
by means of generalized deformed oscillator respectively, and 
Quesne \cite{quesne1} related it to generalized deformed parafermion.
Junker {\it et al.} \cite{jr} constructed (nonlinear) coherent states of ${\cal H}$ 
for the conditionally exactly solvable model with the radial potential of harmonic 
oscillator, and Sunilkumar {\it et al.} \cite{sbjps} did for the quantum optical 
model of four-photon process governed by quadrilinear boson Hamiltonian. 
Recently, Beckers {\it et al.} \cite{bbd} discussed single-variable differential 
realizations of ${\cal H}$ and gave a unitary two-boson realization. 
Ruan {\it et al.} \cite{rjr} studied indecomposable representations of 
the PAMA with a quadratic term, and then from these representations obtained 
its inhomogeneous one-, two- and three-boson realizations. 
In the present work we shall study in detail for ${\cal H}$ various single-mode 
realizations of importance in physical applications, which are 
generalizations of Holstein-Primakoff realization, \cite{hp-1} Dyson realization 
\cite{dyson} and Villain realization \cite{villain} for SU(2).

Let us begin with reviewing briefly the elementary results of boson operators \cite{bl}:
$a^+$ and $a$ ($a$ is adjoint to $a^+$, i.e., $a = (a^+)^\dagger$, 
$a^+ = (a)^\dagger$) are creation and 
annihilation boson operators respectively, which, together with particle number 
operator $\hat{n}$ $\equiv$ $a^+ a$, satisfy commutation relations
\begin{eqnarray}
    [ a, \; a^+ ] = 1,  \hspace{3mm}
    [ \hat{n}, \mbox{} a^+ ] = a^+  , \hspace{3mm}
    [ \hat{n}, \mbox{} a ] = - a.
\label{an-cr}
\end{eqnarray}
Furthermore, the complete set of basis vectors of Fock space,
\{$\left| n \right \rangle$; $n=0$, 1, 2, ...\}, may be constructed from  
$\left| 0 \right \rangle$ by using the definition
\begin{eqnarray}
   \left| n \right \rangle = \frac{ (a^+)^n }{ \sqrt{n!} } \left| 0 \right \rangle.
\label{state}
\end{eqnarray}
In fact, these vectors are normalized eigenvectors of $\hat{n}$ belonging to 
eigenvalue $n$
\begin{equation}
  \hat{n} \left| n \right \rangle = n \left| n \right \rangle, 
\label{n-eigen}
\end{equation}
and satisfy
\begin{eqnarray}
	a^+ \left| n \right \rangle =  \sqrt{n+1} \left| n+1 \right \rangle, \hspace{5mm}
	a \left| n \right \rangle =  \sqrt{n} \left| n-1 \right \rangle.
\label{aa+n}
\end{eqnarray}

Now turn to the Higgs algebra ${\cal H}$, whose three elements \{$J_+$, $J_-$, $J_3$\} 
satisfy the following commutation relations
\begin{equation}
	[ J_+, \hspace{1mm} J_-] = C_1 J_3 + C_3 J^3_3, \hspace{4mm} 
	[ J_3, \hspace{1mm} J_{\pm} ] = \pm J_{\pm} ,    
\label{higgs}
\end{equation}
where $C_1$ and $C_3$ are arbitrary real numbers. When $C_1=2$ (or $-2$) and $C_3=0$, 
${\cal H}$ defined by Eq. (\ref{higgs}) goes back to SU(2) (or its non-compact type 
SU(1,1)). Similar to the Casimir invariant of SU(2), \cite{bl} the invariant of
${\cal H}$ reads 
\begin{equation}
\begin{array}{rl}
{\cal C} & = J_{+} J_{-} + J_{-} J_{+} + ( C_1 + \frac12 C_3 ) J_3^2 
             + \frac12 C_3 J_3^4    \\
{}       & = 2 J_{-} J_{+} + C_1 J_3 + ( C_1 + \frac12 C_3 ) J_3^2 
             + C_3 J_3^3 + \frac12 C_3 J_3^4, 
\end{array}
\end{equation}
which commutes with the generators of ${\cal H}$, 
i.e., 
\begin{equation}
 [{\cal C}, \mbox{} J_{+} ] = 
 [{\cal C}, \mbox{} J_{-} ] = 
 [{\cal C}, \mbox{} J_3 ] = 0.
\end{equation}
It is worthy of reminding the readers that the constant $C_1$ in Eq. (\ref{higgs})
is remained for convenience though it may become some fixed real number, say $q$, 
by rescaling the elements $J_{\pm}$:
$J_{\pm} \rightarrow \sqrt{\frac{q}{C_1}} J_{\pm}$.

Making use of the very parallel treatment of angular momentum in quantum mechanics, 
\cite{bl} it is not difficult to obtain the following unitary representation of ${\cal H}$ 
in the common eigenvectors
$\left| jm \right \rangle$ of the elements \{${\cal C}$, $J_3$\}, with $j$ and $m$ 
labeling the eigenvalues of ${\cal C}$ and $J_3$ respectively, as \cite{bbd,ruan}
\begin{equation}
\begin{array}{rl}
\left \langle j m+1 \right | J_+ \left| jm \right \rangle =  
    &  \sqrt{ \frac12 C_1[j(j+1)- m(m+1)] + \frac14 C_3[j^2 (j+1)^2 - m^2 (m+1)^2 ]}, \\   
\left \langle j m-1 \right | J_- \left| jm \right \rangle =  
    &  \sqrt{ \frac12 C_1[j(j+1)- m(m-1)] + \frac14 C_3[j^2 (j+1)^2 - m^2 (m-1)^2 ]}, \\   
\left \langle j m \right | J_3 \left| jm \right \rangle =  
    &  m, 	\\   
\left \langle j m \right | {\cal C} \left| jm \right \rangle =  
    &  C_1 j (j+1) + \frac12 C_3 j^2 (j+1)^2.
\label{urep}
\end{array}
\end{equation}
Here we have adopted the similar phase factor as the Condon-Shortley rule
of SU(2) so that the matrix elements of $J_{\pm}$ given by Eq. (\ref{urep}) 
are real. In Eq. (\ref{urep}), $j$ may take half-integers, 0, 1/2, 1, 3/2,..., 
and for the finite dimensional representation with a fixed $j$, the values that 
$m$ may take, being a part of \{$-j$, $-j+ 1$,..., $j$\}, are different 
for the different $C_1$ and $C_3$. \cite{ruan}

Mapping the sequence of half-integers $m$ onto a set of non-negative
integers $n$, by the displacement
\begin{equation}
\begin{array}{rl}
	m = j - n,
\label{disp}
\end{array}
\end{equation}
it follows that Eq. (\ref{urep}) becomes, in a convenient notation with omitting 
the eigenvalue $j$,
\begin{equation}
\begin{array}{rl}
\left \langle n-1 \right | J_+ \left| n \right \rangle =  
    &  \frac12 \sqrt{n (2j -n +1) \{ 2 C_1 + C_3 [ 2 j^2 - (n -1) (2j -n) ]\} }, 	\\   
\left \langle n+1 \right | J_- \left| n \right \rangle =  
    &  \frac12 \sqrt{(n+1) (2j -n ) \{ 2 C_1 + C_3 [ 2 j^2 - n (2j -n - 1) ]\} }, 	\\   
\left \langle n \right | J_3 \left| n \right \rangle =  &  j-n.
\label{urep-n}
\end{array}
\end{equation}
Comparing Eq. (\ref{urep-n}) with Eq. (\ref{aa+n}), we notice that the behavior of 
$J_+$ ($J_-$) acting on the basis vector $\left| n \right \rangle$ to produce another 
basis vector $\left| n-1 \right \rangle$ ($\left| n+1 \right \rangle$) is similar to 
that of the boson operator $a$ ($a^+$). 
Using Eqs. (\ref{an-cr})-(\ref{aa+n}), we can obtain from 
Eq. (\ref{urep-n}) a single-mode realization of ${\cal H}$
\begin{equation}
\begin{array}{rl}
B_1^{\mbox{HP}} (J_+) = & \frac{1}{2} \sqrt{(2j - \hat{n}) \left\{ 2 C_1 + C_3
	[ 2 j^2 - (2 j - \hat{n} -1) \hat{n}] \right\} } a, \\
B_1^{\mbox{HP}} (J_-) = & \frac{1}{2} a^+ \sqrt{(2j - \hat{n}) \left\{ 2 C_1 + C_3 
	[ 2 j^2 - (2 j - \hat{n} -1) \hat{n}] \right\} }  ,\\
B_1^{\mbox{HP}} (J_3) = & j - \hat{n}.
\label{hp-1}
\end{array}
\end{equation}
This realization (\ref{hp-1}) preserves the unitary relation 
$ B_1^{\mbox{HP}}(J_+) = \left( B_1^{\mbox{HP}} (J_-) \right)^{\dagger} $. 
Taking $C_1 = 2$ and $C_3 = 0$ in Eq. (\ref{hp-1}) leads to 
the standard Holstein-Primakoff realization of SU(2), \cite{hp-1} which, 
together with the Villain realization to be discussed later, is frequently 
applied to solving various antiferromagnetic and ferromagnetic models.
It can be seen from Eq. (\ref{urep-n}) that the space that 
the operators $B_1^{\mbox{HP}} (J_{\mu})$ ($\mu = \pm$, 3) act on 
is a subspace of the boson Fock space \{$\left| n \right \rangle$\} 
with $n$ limited in order that the value of the equation in 
the square-root operator of Eq. (\ref{urep-n}) must be greater than 
or equal to zero, for example, if $Z_- \leq Z_+ \leq 2j $, where 
\begin{equation}
	Z_{\pm} =  j - \frac12 \pm \frac12 \sqrt{2 -(2j+1)^2 - 8 C_1 / C_3},
\label{z+-}
\end{equation}
then $n$ may take $\{Z_+ \} \leq n \leq 2j$ or $n \leq [Z_-]$, where
for a real number $x$ the symbol $[x]$ means taking integral part of $x$,
whereas $\{x \}$ does taking a integer greater than $x$.

The form of Eq. (\ref{hp-1}) reminds us that the boson realizations of $J_+$ 
and $J_-$ can be chosen in the more general form 
\begin{equation}
    B_k (J_+) = f_1 (\hat{n}) a^k, \hspace{5mm}  B_k (J_-) = (a ^+)^k f_2 (\hat{n}),  
\label{jj-equ}
\end{equation}
where $k$ is a positive integer, $f_1 (\hat{n})$ and $f_2 (\hat{n})$ are the functions
of $\hat{n}$ merely, whereas the boson realization of $J_3$ holds the same expression 
as $B_1^{\mbox{HP}}(J_3)$, i.e., $B_k (J_3)= j - \hat{n}$, which is Hermitian. 
Here for $k=1$, 2, 3,..., we call $B_k (J_{\mu})$ ($\mu = \pm$, 3) the realizations of 
simple type, quadratic type, cubic type and so on respectively owing to the fact that 
the action of $B_k (J_{\pm})$ on the basis vector $\left| n \right \rangle$ of 
the boson Fock space will give another basis vector $\left| n' \right \rangle$ with 
$n' = n \mp k$.
The first commutation relation of Eq. (\ref{higgs}) requires that $f_1 (\hat{n})$ and 
$f_2 (\hat{n})$ satisfy the following difference equation
\begin{equation}
\begin{array}{l}
    (\hat{n}+ 1) \prod\limits_{i=2}^{k} (\hat{n}+ 2^{i-2} +1) f_1 (\hat{n}) f_2 (\hat{n}) 
	- \prod\limits_{i=1}^{k} (\hat{n} - i +1) f_1(\hat{n} - k) f_2(\hat{n} - k) \\
	= C_1 (j - \hat{n}) + C_3 (j - \hat{n})^3,
\label{fg-equ}
\end{array}
\end{equation}
with the help of the relatios
\begin{equation}
\begin{array}{rl}
	(a^+)^k f_i (\hat{n}) = & f_i (\hat{n} - k) (a^+)^k, \\
	a^k f_i (\hat{n})   = & f_i (\hat{n} + k) a^k,   
	\hspace{5mm} i = 1, \: 2.
\end{array}
\end{equation}
For a given $k$, the solution of Eq. (\ref{fg-equ}) with $k$ initial conditions
\begin{equation}
	\left\{ f_1 (l) f_2 (l) = C_1 (j - l) + C_3 (j - l)^3; \;\; 
	l = 0, \, 1, ..., k-1 \right\}
\end{equation}
is unique, for example,
\begin{equation}
	f_1(\hat{n}) f_2(\hat{n}) 
	= \frac{1}{4} (2j - \hat{n}) 
	\left\{ 2 C_1 + C_3 [ 2 j^2 - (2 j - \hat{n} -1) \hat{n}] \right\}
\label{solu-1}
\end{equation}
for $k = 1$, and
\begin{equation}
\begin{array}{rl}
	f_1(\hat{n}) f_2(\hat{n})  = & \frac{1}{16 (\hat{n} + 1) (\hat{n} + 2)} 
		\left\{ 2 C_1 [ (-1)^{\hat{n}} (2 j + 1) - (\hat{n} + 1)  \right.    \\
	{} &	+ (2 j - \hat{n}) (2 \hat{n} + 3)] + C_3 [ 1 + 6 j^2 (2 j -1)  \\
	{} &	+ (-1)^{\hat{n}} (2 j + 1) (2 j^2 + 2 j -1 )   \\
	{} &	+ \left. 2 \hat{n} (2 j - \hat{n} -2 )
		(2 j^2 - (2 j - \hat{n}) (\hat{n} + 2))]  \right\}
\label{solu-2}
\end{array}
\end{equation}
for $k = 2$.

The above solutions show that we may have some freedom in the choice of the functions 
$f_1 (\hat{n})$ and $f_2 (\hat{n})$. 

(1) If the unitary relation $ B_k (J_+) = (B_k (J_-))^{\dagger} $ need satisfying, that is, 
$f_1 (\hat{n}) = f_2 (\hat{n})$, then solving Eqs. (\ref{solu-1}) and (\ref{solu-2})
and substituting them into Eq. (\ref{jj-equ}) respectively, we may regain 
$B_1^{\mbox{HP}} (J_{\mu})$ (see Eq. (\ref{hp-1})) for $k=1$, and obtain
\begin{equation}
\begin{array}{rl}
B_2^{\mbox{HP}} (J_+) = & \frac{1}{4} [(\hat{n} + 1) (\hat{n} + 2)]^{-1/2}
	\left\{ 2 C_1 [ (-1)^{\hat{n}} (2 j + 1) - (\hat{n} + 1)           \right.   \\
	{} &	+ (2 j - \hat{n}) (2 \hat{n} + 3)]  + C_3 [ 1 + 6 j^2 (2 j -1)       \\
	{} &	+ (-1)^{\hat{n}} (2 j + 1) (2 j^2 + 2 j -1 )                     \\
	{} &	+ \left. 2 \hat{n} (2 j - \hat{n} -2 )
	(2 j^2 - (2 j - \hat{n}) (\hat{n} + 2))]  \right\} ^{1/2} a^2,    \\
B_2^{\mbox{HP}} (J_-)  = & \frac{1}{4} (a^+)^2 [(\hat{n} + 1) (\hat{n} + 2)]^{-1/2}
	\left\{ 2 C_1 [ (-1)^{\hat{n}} (2 j + 1) - (\hat{n} + 1)           \right.   \\  
	{} &	+ (2 j - \hat{n}) (2 \hat{n} + 3)]  + C_3 [ 1 + 6 j^2 (2 j -1)       \\
	{} &	+ (-1)^{\hat{n}} (2 j + 1) (2 j^2 + 2 j -1 )                     \\
	{} &	+ \left. 2 \hat{n} (2 j - \hat{n} -2 )(2 j^2 - (2 j - \hat{n}) (\hat{n} + 2))] 
	\right\} ^{1/2},   \\
B_2^{\mbox{HP}} (J_3)  = & j - \hat{n}
\label{hp-2}
\end{array}
\end{equation}
for $k=2$. Similar to $B_1^{\mbox{HP}} (J_{\mu})$, the values of $n$ in the matrix 
elements of $B_2^{\mbox{HP}} (J_{\mu})$ in the Fock space need limiting as well.
We call Eqs. (\ref{hp-1}) and (\ref{hp-2}) the Holstein-Primakoff-like 
realizations of simple type and quadratic type of ${\cal H}$ respectively.

(2) If the unitary relation need not satisfying, it follows from Eqs. (\ref{solu-1}) 
and (\ref{solu-2}) that the conventional choice $f_2(\hat{n}) = 1$ 
(or $f_1(\hat{n}) = 1$) may immediately give rise to another kind of single-mode 
realization of importance
\begin{equation}
\begin{array}{rl}
B_1^{\mbox{D}} (J_+)  = & \frac{1}{4} (2j - \hat{n}) \{ 2 C_1 + C_3
	[ 2 j^2 - (2 j - \hat{n} -1) \hat{n}]  \} a,   \\
B_1^{\mbox{D}} (J_-)  = & a^+ , \\
B_1^{\mbox{D}} (J_3)  = & j - \hat{n}  
\label{dyson-1}
\end{array}
\end{equation}
for $k=1$, which may also be obtained by means of the approach adopted in 
Ref. \cite{rjr}, and
\begin{equation}
\begin{array}{rl}
B_2^{\mbox{D}} (J_+)  = & \frac{1}{16 (\hat{n} + 1) (\hat{n} + 2)} 
	\left\{ 2 C_1 [ (-1)^{\hat{n}} (2 j + 1) - (\hat{n} + 1) 
	+ (2 j - \hat{n}) (2 \hat{n} + 3)] \right.  \\
	{} &	+ C_3 [ 1 + 6 j^2 (2 j -1) 
		+ (-1)^{\hat{n}} (2 j + 1) (2 j^2 + 2 j -1 )   \\
	{} &	+ \left. 2 \hat{n} (2 j - \hat{n} -2 )
		(2 j^2 - (2 j - \hat{n}) (\hat{n} + 2))]  \right\} a^2,  \\
B_2^{\mbox{D}} (J_-) = & (a^+)^2 , \\
B_2^{\mbox{D}} (J_3) = & j - \hat{n} 
\label{dyson-2}
\end{array}
\end{equation}
for $k=2$. When $C_1=2$ and $C_3=0$, Eq. (\ref{dyson-1}) becomes the standard Dyson 
realization of SU(2) introduced originally by Dyson \cite{dyson} in his study of 
spin-wave interactions, hence, we call Eqs. (\ref{dyson-1}) and (\ref{dyson-2}) 
the Dyson-like realizations of simple type and quadratic type of ${\cal H}$ respectively.
Different from the Holstein-Primakoff-like realizations, (\ref{hp-1}) and (\ref{hp-2}), 
no square-root operator appearing in the Dyson-like realizations, (\ref{dyson-1}) and 
(\ref{dyson-2}), may not only avoid the convergence questions associated with the expansion 
of square-root operator but also make the value of $n$ in \{$\left| n \right \rangle$\} 
unlimited, i.e., $n=0$, 1, 2,....

Furthermore, it is not difficult to find that the non-unitary Dyson-like realizations 
$B_i^{\mbox{D}} (J_{\mu})$ ($i=1$, 2) may be related to the unitary Holstein-Primakoff-like
realizations $B_i^{\mbox{HP}} (J_{\mu})$ by their corresponding similarity transformations $S_i$
\begin{equation}
\begin{array}{rl}
	S_i B_i^{\mbox{D}} (J_3) S_i^{-1} = & B_i^{\mbox{HP}} (J_3) = B_i^{\mbox{D}} (J_3), \\
	S_i B_i^{\mbox{D}} (J_{\pm}) S_i^{-1} = & B_i^{\mbox{HP}} (J_{\pm}), \hspace{4mm}
	i=1, \; 2.
\label{s-t}
\end{array}
\end{equation}
Considering the unitray relations 
$ B_i^{\mbox{HP}}(J_+) = \left( B_i^{\mbox{HP}} (J_-) \right)^{\dagger} $, 
we obtain from the second equation of Eq. (\ref{s-t})
\begin{equation}
	U_i^{-1} \left(B_i^{\mbox{D}} (J_-) \right)^{\dagger} U_i = B_i^{\mbox{D}} (J_+),
	\hspace{5mm}  i=1, \; 2 ,
\label{u-t}
\end{equation}
where $ U_i \equiv S_i^{\dagger} S_i $ are Hermitian operators. 
Note that $B_i^{\mbox{D}}(J_3)$ are already Hermitian. The equations (\ref{u-t}) 
are called unitarization of the Dyson-like realizations. As an example, let us calculate 
concretely the explicit expression of $S_1$.
The first equation of Eq. (\ref{s-t}) implies that $S_1$ commutes with $J_3$ and is 
at most the function of $\hat{n}$, thus, calculating the matrix element of Eq. (\ref{u-t}) 
between the basis vectors $\left \langle n-1 \right |$ and $\left| n \right \rangle$ 
and using the former two equations of Eq. (\ref{dyson-1}), we derive
\begin{equation}
	\left \langle n \right | S_1 \left| n \right \rangle ^2 = 
	\frac{1}{4} (2j - n +1) \{ 2 C_1 + C_3
	[ 2 j^2 - (2 j - n) (n-1) ] \}
	\left \langle n - 1 \right | S_1 \left| n - 1 \right \rangle ^2.
\label{s-eq}
\end{equation}
Solving Eq. (\ref{s-eq}) with the initial condition $\langle 0 | S_1 | 0 \rangle = q$
($q$ is a real number),
and then using Eq. (\ref{n-eigen}), we obtain 
\begin{equation}
	S_1 = \pm \sqrt{q (- C_3 / 4)^{\hat{n}} (- 2 j)_{\hat{n}}
		(- Z_+)_{\hat{n}} (- Z_-)_{\hat{n}} } ,
\label{s-solu}
\end{equation}
where $Z_{\pm}$ have been given by Eq. (\ref{z+-}), 
and $(q)_{\hat{n}}$ is a operator function of $\hat{n}$ for a fixed real number $q$
\begin{equation}
	(q)_{\hat{n}} = q (q +1) ... (q + \hat{n} -1 ),
\end{equation}
whose expectation value in the boson Fock space \{$\left| n \right \rangle$\} is in fact 
the usual Pochhammer symbol $(q)_{n}$ for a positive integer $n$, i.e.,
$
	\left \langle n \right | (q)_{\hat{n}}  \left| n \right \rangle \equiv 
	(q)_n = q (q +1) ... (q + n -1 ),
$
with setting $\left \langle 0 \right | (q)_{\hat{n}}  \left| 0 \right \rangle = 1$.

Finally consider the third kind of single-mode realization, which is based upon 
the Villain-like realizations of ${\cal H}$ in terms of a coordinate $X$ and 
the corresponding momentum operator $P= - \mbox{i} \frac{\mbox{d}}{\mbox{d} X}$.
However, the Villain-like realization of ${\cal H}$ is not unique, here we give 
two different realizations
\begin{equation}
\begin{array}{rl}
V_1 (J_+) = & \exp(\mbox{i} X) 
	\sqrt{g_1^2 - \frac14 C_3 [ P (P+1) ]^2 - \frac12 C_1 (P + \frac12)^2}, \\
V_1 (J_-) = & \sqrt{g_1^2 - \frac14 C_3 [ P (P+1) ]^2 - \frac12 C_1 (P + \frac12)^2}
	\exp(- \mbox{i} X),\\
V_1 (J_3) = & P,
\label{v-1}
\end{array}
\end{equation}
and
\begin{equation}
\begin{array}{rl}
V_2 (J_+) = & \frac{1}{2 \sqrt{C_3}} \exp(\mbox{i} X) 
	\sqrt{g_2^2 - (C_3 P^2 +  C_3 P + C_1)^2}, \\
V_2 (J_-) = & \frac{1}{2 \sqrt{C_3}} 
	\sqrt{g_2^2 - (C_3 P^2 +  C_3 P + C_1)^2} \exp(- \mbox{i} X),\\
V_2 (J_3) = & P,
\label{v-2}
\end{array}
\end{equation}
with $C_3 \not= 0$, where $g_1$ and $g_2$ are the functions of the eigenvalue $j$, 
their explicit expressions may be respectively chosen as
\begin{equation}
\begin{array}{rl}
	g_1 = & \pm \sqrt{\frac12 C_1 (j + \frac12 )^2 + \frac14 C_3 j^2 (j + 1)^2}, \\
	g_2 = & \pm \frac12 \sqrt{\frac12 C_1^2 + C_1 C_3 j (j+1) 
		+ \frac12 C_3^2 j^2 (j + 1)^2}
\end{array}
\end{equation}
in order that the corresponding Casimir operators, $V_1 ({\cal C})$ and $V_2 ({\cal C})$,
have the same value as $\left \langle j m \right | {\cal C} \left| jm \right \rangle$, 
i.e.,
\begin{equation}
     V_1 ({\cal C}) = V_2 ({\cal C}) = C_1 j (j+1) + \frac12 C_3 j^2 (j+1)^2.
\end{equation}
By means of the equation 
\begin{equation}
   \exp (\mbox{i} \theta X) P^k \exp ( - \mbox{i}  \theta X) = (P- \theta)^k,
\end{equation}
where $\theta$ is a formal parameter, it is easy to check out the validity of 
the commutation relations (\ref{higgs}) for the Villain-like realizations 
(\ref{v-1}) and (\ref{v-2}).
Taking $C_1=2$ and $C_3=0$, the first Villain-like realization, Eq. (\ref{v-1}), 
becomes the standard Villain realization of SU(2) \cite{villain}
with the correspondence $X \leftrightarrow \varphi_{\bf R}$, 
$P \leftrightarrow S^z_{\bf R}$ and $g_1 = s + \frac12$.

Thus, substituting 
\begin{equation}
X = \frac{1}{\sqrt{2}}( a + a^+ ), \hspace{3mm} 
P = - \frac{\mbox{i}}{\sqrt{2}} ( a - a^+ ),
\label{xp-aa}
\end{equation}
into Eqs. (\ref{v-1}) and (\ref{v-2}) respectively, we may obtain the following 
two single-mode realizations of ${\cal H}$
{\footnotesize
\begin{equation}
\begin{array}{rl}
B_1^{\mbox{V}}(J_+) = & \exp[ \frac{\mbox{i}}{\sqrt{2}} (a + a^+)] 
  \sqrt{g_1^2 - \frac14 C_3 [ \frac12 (a - a^+)^2 + \frac{\mbox{i}}{\sqrt{2}} (a - a^+)]^2
  - \frac12 C_1 [\frac{\mbox{i}}{\sqrt{2}} (a - a^+) - \frac12 ]^2 }, \\
B_1^{\mbox{V}}(J_-) = & 
  \sqrt{g_1^2 - \frac14 C_3 [ \frac12 (a - a^+)^2 + \frac{\mbox{i}}{\sqrt{2}} (a - a^+)]^2
  - \frac12 C_1 [\frac{\mbox{i}}{\sqrt{2}} (a - a^+) - \frac12 ]^2 }
  \exp[ - \frac{\mbox{i}}{\sqrt{2}} (a + a^+)], \\
B_1^{\mbox{V}}(J_3) = & - \frac{\mbox{i}}{\sqrt2} (a - a^+),
\label{v-boson-1}
\end{array}
\end{equation}
}
and
\begin{equation}
\begin{array}{rl}
B_2^{\mbox{V}}(J_+) = & \frac{1}{2 \sqrt{C_3}}\exp[ \frac{\mbox{i}}{\sqrt{2}} (a + a^+)] 
	\sqrt{g_2^2 - [ \frac{1}{2} C_3 (a - a^+)^2 + \frac{\mbox{i}}{\sqrt{2}} 
	C_3 (a - a^+ ) - C_1 ]^2}, \\
B_2^{\mbox{V}}(J_-) = & \frac{1}{2 \sqrt{C_3}} 
	\sqrt{g_2^2 - [ \frac{1}{2} C_3 (a - a^+)^2 + \frac{\mbox{i}}{\sqrt{2}} 
	C_3 (a - a^+ ) - C_1 ]^2}
	\exp[ - \frac{\mbox{i}}{\sqrt{2}} (a + a^+)], \\
B_2^{\mbox{V}}(J_3) = & - \frac{\mbox{i}}{\sqrt2} (a - a^+).
\label{v-boson-2}
\end{array}
\end{equation}
Obviously, the above two realizations are unitary. 

In summary, we have obtained for the Higgs algebra the explicit expressions for 
the unitary Holstein-Primakoff-like realizations of simple and quadratic types, 
the non-unitary Dyson-like realizations of simple and quadratic types, and 
two unitary realizations based upon the Villain-like realizations, some of which 
can be found their prototypes for SU(2).
It can be checked that all these single-mode realizations, 
(\ref{hp-1}), (\ref{hp-2}), (\ref{dyson-1}), (\ref{dyson-2}), (\ref{v-boson-1})
and (\ref{v-boson-2}), 
satisfy the commutation relations (\ref{higgs}) of the Higgs algebra. Furthermore, 
we have revealed the fact that the Holstein-Primakoff-like realizations and 
the Dyson-like realizations of the Higgs algebra may be related by the corresponding 
similarity transformations.
Due to the tight relations between boson operators and differential 
operators \cite{bl}, for another example except for Eq. (\ref{xp-aa}), 
$a$ $\leftrightarrow$ $\frac{\mbox{d}}{\mbox{d} X}$ and $a^+$ $\leftrightarrow$ $X$,
the differential realizations of the Higgs algebra may be obtained directly from 
the above various single-mode realizations. They are not given here. 
These results may be applied to some typical quantum mechanical systems characterized by 
the Higgs algebra to obtain the corresponding energy spectra, (nonlinear) coherent states 
and so on.
The method adopted in this paper may be used to discuss multi-mode realizations of 
the Higgs algebra and to treat the more general PAMA, for example, with the commutator 
$ [ J_+, \hspace{1mm} J_-] = \sum_{i=0}^{n} J_3^{i}$. 
These studies are now under way.

\vspace{5mm}

This work is supported by National Natural Science Foundation of China
(19905005), Major State Basic Research Development Program (G2000077400 and
G2000077604) and Tsinghua Natural Science Foundation (985 Program).


\newpage

\begin{thebibliography}{99}
\bibitem{rocek} Ro\v{c}ek M 1991 {\it Phys. Lett.} {\bf B255} 554
\bibitem{bl} Biedenharn L C and Louck J D 1981
	{\it Angular Momentum in Quantum Physics}
	(Massachusetts: Addison-Wesley) 
\bibitem{higgs} Higgs P W 1979 {\it J. Phys.} {\bf A12} 309 
\bibitem{zhedanov} Zhedanov A S 1991 {\it Mod. Phys. Lett.} {\bf A7} 507
\bibitem{jimbo} Jimbo M 1985 {\it Lett. Math. Phys.} {\bf 10} 63
\bibitem{dask} Daskaloyannis C 1991 {\it J. Phys.} {\bf A24} L789  
\bibitem{bdk} Bonatsos D, Daskaloyannis C and Kolokotronis P 1994 
	{\it Phys. Rev.} {\bf A50} 3700
\bibitem{bkd} Bonatsos B, Kolokotronis P and Daskaloyannis C 1995 
	{\it Mod. Phys. Lett.} {\bf A10} 2197 
\bibitem{quesne1} Quesne Q 1994 {\it Phys. Lett.} {\bf A193} 245 
\bibitem{jr} Junker G and Roy P 1999 {\it Phys. Lett.} {\bf A257} 113
\bibitem{sbjps} Sunilkumar V, Bambah B A, Jagannathaan R, Panigrah P K
	and Srinivasan V 2000 {\it J. Opt.} {\bf B2} 126
\bibitem{bbd} Beckers J,  Brihaye Y and Debergh N 1999 {\it J. Phys.} {\bf A32} 2791
\bibitem{rjr} Ruan D, Jia Y F and Ruan W 2000 {\it J. Math. Phys.} {\bf 42} 2718
\bibitem{hp-1} Holstein T and Primakoff H 1940 {\it Phys. Rev.} {\bf 58} 1098.
\bibitem{dyson} Dyson J F 1956 {\it Phys. Rev.} {\bf 102} 1217
\bibitem{villain} Villain J 1974 {\it J. de Phys.} {\bf 35} 27
\bibitem{ruan} Ruan D 2001 {\it Frontiers in Quantum Mechanics}
	edited by Zeng J Y, Pei S Y and Long G L (Beijing: Beijing University)    
\end{thebibliography}
\end{document}